\documentclass{ws-ijmpa}

\begin{document}

\newcommand\be{\begin{equation}}
\newcommand\ee{\end{equation}}
\newcommand\bea{\begin{eqnarray}}
\newcommand\eea{\end{eqnarray}}
\newcommand\bseq{\begin{subequations}} 
\newcommand\eseq{\end{subequations}}
\newcommand\bcas{\begin{cases}}
\newcommand\ecas{\end{cases}}
\newcommand{\p}{\partial}
\newcommand{\f}{\frac}


%
%

\title{Cosmological implications of an evolutionary quantum gravity}

\author{Marco Valerio Battisti}

\address{ICRA - International Center for Relativistic Astrophysics\\Phys. Dept. (G9), University of Rome ``La Sapienza'' P.le A. Moro 5, 00185 Rome, Italy\\
battisti@icra.it}

\author{Giovanni Montani}

\address{ICRA - International Center for Relativistic Astrophysics\\Phys. Dept. (G9), University of Rome ``La Sapienza'' P.le A. Moro 5, 00185 Rome, Italy\\ENEA C.R. Frascati (Dipartimento F.P.N.), Via Enrico Fermi 45, 00044 Frascati, Rome, Italy\\ICRANET C.C. Pescara, P.le della Repubblica 10, 65100 Pescara, Italy\\
montani@icra.it}

\maketitle

\begin{abstract}
The cosmological implications of an evolutionary quantum gravity are analyzed in the context of a generic inhomogeneous model. The Schr\"{o}dinger problem is formulated and solved in the presence of a scalar field, an ultrarelativistic matter and a perfect gas regarded as the dust-clock. Considering the actual phenomenology, it is shown how the evolutionary approach overlaps the Wheeler-DeWitt one. 

\end{abstract}

\ccode{PACS numbers: 04.20.Dw, 98.80.Qc}

\bigskip

The generally covariant system ``per excellence'' is the gravitational field in General Relativity (GR), i.e. it is invariant under arbitrary changes of the space-time coordinates (4-dimensional diffeomorphisms). In particular, the Hamiltonian of such a theory is a combination of constraints, i.e. is (weakly) zero and one treats on the same footing observables and constants of motion. This non trivial feature of GR is reflected at quantum level by the so-called {\it problem of time} [\refcite{protim}] where the Schr\"{o}dinger equation is replaced by a Wheeler-DeWitt one, in which the time coordinate disappears from the formalism. 

In order to overcame this behavior, the most suitable way is to consider a matter field to define a relational time, i.e. evolve the gravitational field with respect to another one. In particular, an useful choice for the matter field resides in an incoherent dust, i.e. a dust which interacts only gravitationally [\refcite{Mo02}] (see also [\refcite{KuBr}]). In fact, it is possible to show how there exist a dualism between time evolution and the presence of a dust fluid in quantum gravity [\refcite{MeMo04},\refcite{BaMo06}]. Indeed, starting from a Schr\"{o}dinger equation for the gravitational field (described by $\Psi=\Psi(t,\left\{h_{ij}\right\})$ on the super-space of the $3$-geometries $\left\{h_{ij}\right\}$)
\be\label{eipro}
i\p_t \Psi=\hat{\mathcal{H}}\Psi\equiv\int_{\Sigma}d^3x\left(N\hat{H}\right)\Psi \quad \Rightarrow \quad \hat{H}\chi=\epsilon\chi,\quad \hat{H_i}\chi=0,
\ee 
$H$ being the super-Hamiltonian, $N=N(t)$ the lapse function and $\Sigma$ the compact Cauchy surface, and taking the zero-order WKB limit for the (stationary) wave function of the Universe: $\chi\sim e^{iS}$, a new matter contribution in the system dynamics appears whose energy density is $\rho=-{\epsilon(x)}/{2\sqrt h}$. Therefore, a dust fluid co-moving with the slicing Cauchy surfaces, i.e. with $T_{\mu\nu}=\rho n_\mu n_\nu$, is induced in the theory. On the other hand, if we include {\it ab initio} a dust into the Universe dynamics foliating the space-time manifold, we obtain the same eigenvalues problem (\ref{eipro}) as above. This way, a dust fluid is a good choice to realize a clock in quantum gravity.

From such a point of view, this matter contribution has to emerge in any system which undergoes a classical limit and therefore it must concern the history of the Universe. Since it is expected the quantum behavior of the Universe takes place in the Planck era, only Planck mass particles can be reliably inferred and appropriately described by a perfect gas contribution. In particular, we construct a solution in which such a term is included into the dynamics and we link the quantum number, associated to its energy density, to the eigenvalue $\epsilon$. In doing that, we use the Planck mass particles as a clock for the quantum dynamics and, at the same level, we induce by them a non-relativistic matter component into the early Universe. Of course, the phenomenological implications of such an approach have to be analyzed in order to comprehend if this matter field (related to the the non-zero eigenvalue of the super-Hamiltonian) is observable or not in the actual Universe. As we will see, no phenomenology can came out today from our dust fluid (see [\refcite{BaMo06}]).

In quantum gravity, the early Universe has to be described by a generic inhomogeneous model [\refcite{BKL82}] since no symmetries can be assumed in the model because of the (expected) quantum fluctuation of the metric structure concerning the space-time manifold. Such a model is classically described by the line element [\refcite{BeMo04}]
\begin{equation}\label{metrten}%
ds^2=N^2dt^2-h_{ij}(dx^i+N^idt)(dx^j+N^jdt),
\end{equation}
the spatial metric $h_{ij}$ being $h_{ij}=e^{q_a}\delta_{ad}O^a_b O^d_c \partial_i y^b \partial_j y^c$, while $q_a$ and $y^a$ are the dynamical variables and $O_b^a(x)$ are $SO(3)$ matrices. In particular, the $y^a$ variables can be used as new spatial variables in order to identically solve the super-momentum constraint $H_i(x)\approx0$. This way, all the dynamical information about this model are contained in the scalar constraint $H(x)\approx0$. As we said, we add a perfect gas contribution into the gravitational dynamics in order to represent the Planck-mass particles. Furthermore, since we want to describe the early Universe evolution, we have to add also a free scalar field $\phi$ in order to account for the inflaton field and an ultrarelativistic term describing the thermal bath of all fundamental particles (whose mass is negligible with respect to the temperature). Therefore the scalar constraint reads in the Misner variables ($a,\beta_\pm$) [\refcite{Mis69a}]
\be
H(x)=\kappa \left[-\f{p_a^2}a+\f{1}{a^3}\left(p^2_+ +p^2_-\right)\right]+\f{3}{8\pi}\f{p^2_\phi}{a^3}-\f{a^3} {4\kappa l^2_{in}} V(\beta_\pm)+a^3\left(\rho_{ur}+\rho_{pg}\right)\approx0,
\ee 
$\kappa=8\pi l_P^2$ being the Einstein constant, $V(\beta_\pm)$ the potential term which accounts for the spatial curvature of the model and $\rho_{ur}=\mu^2(x)/a^4$ and $\rho_{pg}=\sigma^2(x)/a^5$ are the energy density of the ultrarelativistic and perfect gas contributions respectively. The variables $a$ describes the isotropic expansion while $\beta_\pm$ the shape changes of the Universe and the cosmological singularity appears for $a\rightarrow0$. An important point has now to be stressed. Since the relation $\p_t y^a=N^i\p_i y^a$ holds [\refcite{BeMo04}], in a synchronous reference ($N=1$, $N^i=0$) the variables $y^a$ become passive functions and the dynamics reduces, point by point, to the one of a Bianchi IX model [\refcite{rev}]. This way we can work in the {\it minisuperspace} representation and the field theory is reduced to a 4-dimensional mechanical system.

The evolutionary quantization, summarized in the eigenvalues problem (\ref{eipro}), of this system appears to be (considering the right normal ordering [\refcite{Mo02}])
\be\label{H}%
\left\{\kappa \left[\p_a\f 1 a\p_a-\f 1 {a^3}\left(\p_+^2+\p_-^2\right)\right]-\f{3}{8\pi}\f1{a^3}\p_\phi^2 + -\f{R^3} {4\kappa l^2_{in}}V(\beta_\pm)+\f{\mu^2}a+\f{\sigma^2}{a^2}\right\}\chi= \epsilon\chi,
\ee
where $l_{in}$ denotes the co-moving physical scale of inhomogeneities. The solution of the above problem (for details see [\refcite{BaMo06}]) (\ref{H}) can be obtained neglecting the potential term $V(\beta_\pm)$. In fact, in the region $a^3 \ll \mathcal{O}\left(l_P^2 l_{in}\sqrt{{<K^2>}/|\bar{V}(\beta_{\pm})|}\right)$ such a term can be omitted\footnote{It is possible to show how the quasi-classical limit of this model is reached before the potential term becomes important assuring the self-consistency of our model [\refcite{BaMo06}].} and it is possible to factorize the wave function $\chi(a,\beta_\pm,\phi)$ as $\chi(a,\beta_\pm,\phi)=\theta(a)F(\beta_\pm,\phi)$. Therefore, the functions $F(\beta_\pm,\phi)$ are nothing but plane waves with wave number $K=\sqrt{k^2_\beta+k^2_\phi}$ and $\bar {V}(\beta_{\pm})$ is the average of the potential over the width $\Delta\beta\sim 1/\Delta k_\beta$ where the real wave packets are significantly non-zero, i.e. $\bar {V}(\beta_{\pm})=\int_{\Delta\beta^2}{V(\beta_{\pm})}d^2\beta$. On the other hand, the function $\theta(a)$ satisfies the equation
\be\label{eqR}%
\kappa\f d {da}\left(\f 1 a \f{d\theta}{da} \right) + \left(\kappa \f{K^2}{a^3}+\f{\mu^2}{a}+\f{\sigma^2}{a^2}-\epsilon\right)\theta=0
\ee
and it explicitly reads $\theta(a)=\omega(a)\exp\left[-\f 1 {2l_P^2}\left(a+\epsilon l^2_P/16\pi\right)^2\right]$ where, near the singularity, $\omega(a)=\sum_{n=0}^\infty c_n a^{n+\gamma}$ ($\gamma=1-\sqrt{1-K^2}$) and the coefficients $c_n$ of the series obey a complicated recurrence relation [\refcite{BaMo06}]. Since we require the wave function $\chi(a,\beta_\pm,\phi)$ to decay at large scale factor $a$ (where the potential term becomes important), the series $\omega(a)$ must therefore terminate. This way, we obtain the eigenvalues of our problem (\ref{H}). More precisely, the spectrum
of the Universe super-Hamiltonian $H$ and of the quantum number associated to the ultrarelativistic term are given by the following relations
\be\label{spee}%
\epsilon_{n,\gamma}=\f{\sigma^2}{l_P^2(n+\gamma-1/2)}
\ee
\be\label{mue}%
2(n+\gamma)=\left(\f{l_P\epsilon_{n,\gamma}}{16\pi}\right)^2+\f{\mu^2}{8\pi}.
\ee
It is to be emphasized that the ground state $n=0$ eigenvalue $\epsilon_{0,\gamma }=-\sigma^2/{l_P^2(1/2-\gamma)}$, for $\gamma<1/2$ is negative. Thus the real ground state, according to equation (\ref{mue}) for $\mu^2=0$ ($\gamma \simeq 1.8\cdot 10^{-3}$), $\epsilon_0\simeq-2\sigma^2/{l_P^2}$ is associated to a positive dust energy density. As we can see from (\ref{spee}), the non-zero super-Hamiltonian eigenvalue $\epsilon$, i.e. the evolutionary feature of the model, is linked to the quantum number $\sigma^2$ describing the perfect gas of Planck mass particles. Therefore, as we said, such dust is regarded as the clock for the Universe dynamics and at the same level it appears as a new matter contribution which, in principle, may have reliable implications at cosmological level.

In order to phenomenologically analyze such cosmological implications of our clock, we put by hands a cut-off length in the model. The existence of a fundamental minimal scale has long been expected in quantum gravity and it appears in all the most promising attempts to solve this problem [\refcite{Thi},\refcite{Pol}]. What we do is to require a minimal length $l$ {\it per particle} in the perfect gas, i.e. $l\geq l_P$. This way, the energy density of the perfect gas is constrained to be less then the Planck one ($\rho_{pg}\leq\mathcal{O}(1/{l_P^4})$) and therefore $\sigma^2\leq\mathcal{O}(l_P)$. Therefore, two main conclusions can be inferred. (i) The upper limit for $\sigma ^2$ ensures that {\it the spectrum of the Universe is limited by below} and in particular it appears as $|\epsilon|\le \mathcal{O}(1/l_P)$. (ii) From a phenomenological point of view, {\it the evolutionary approach to quantum gravity overlaps the Wheeler-DeWitt scheme} and therefore it can be inferred as appropriate to describe early stages of the (quantum) Universe without significant traces on the later evolution. In fact, the effects of the new matter to the actual Universe can be easily analyzed by its contribution to the critical parameter, which is $\Omega_{dust}\simeq\rho_{dust}/\rho_{today}\simeq\mathcal{O}( 10^{-60})$. Therefore, the cosmological implications of an evolutionary quantum gravity to the present Universe are completely negligible. For a complete discussion on this subject see [\refcite{BaMo06},\refcite{evo}].


\begin{thebibliography}{0}    

\bibitem{protim}C.J.Isham, {\it Canonical quantum gravity and the problem of time}, (1992), gr-qc/9201011; T.P.Shestakova and C.Simeone, {\it Grav. Cosmol.}, (2004), {\bf 10} 161, gr-qc/0409114; T.P.Shestakova and C.Simeone, {\it Grav. Cosmol.}, (2004), {\bf 10} 257, gr-qc/0409119. 

\bibitem{Mo02}G.Montani, {\it Nucl. Phys. B}, (2002), {\bf 634} 370, gr-qc/0205032.

\bibitem{KuBr}J.D.Brown and K.V.Kuchar, {\it Phys. Rev. D}, (1995), {\bf 51} 5600, gr-qc/9409001.

\bibitem{MeMo04}S.Mercuri and G.Montani, {\it Mod. Phys. Lett. A}, (2004), {\bf 19}, n.20 1519, gr-qc/0312077.

\bibitem{BaMo06}M.V.Battisti and G.Montani, {\it Phys. Lett. B}, (2006), {\bf 637} 203, gr-qc/0604049.

\bibitem{BKL82}V.A.Belinski, I.M.Khalatnikov and E.M.Lifshitz, {\it Adv.\ Phys.}, (1982), {\bf 31} 639.

\bibitem{BeMo04}R.Benini and G.Montani, {\it Phys. Rev. D}, (2004), {\bf 70} 103527, gr-qc/0411044.

\bibitem{Mis69a}C.Misner, {\it Phys. Rev. Lett.}, (1969), {\bf 22} 1071.

\bibitem{rev}For a complete review on this subject see G.Montani, M.V.Battisti, R.Benini and G.Imponente, {\it Int. J. Mod. Phys A}, (2008), in press, arXiv:0712.3008.

\bibitem{Thi} T.Thiemann, {\it Modern Canonical Quantum General Relativity} (2007), CUP Cambridge.

\bibitem{Pol}J.Polchinski, {\it String Theory} (2000), CUP Cambridge. 

\bibitem{evo}M.V.Battisti and G.Montani, {\it Nuovo Cim. B}, (2007), {\bf 122} 179, gr-qc/0701095.

\end{thebibliography}
\end{document}